# The activity of the symbiotic binary Z Andromedae and its latest outburst

J. MERC[1,2], R. GÁLIS[2], M. WOLF[1], L. LEEDJÄRV[3], F. TEYSSIER[4]

(1) Astronomical Institute, Faculty of Mathematics and Physics, Charles University
V Holešovičkách 2, 180 00 Prague, Czech Republic

(2) Institute of Physics, Faculty of Science, P. J. Šafárik University
Park Angelinum 9, 040 01 Košice, Slovak Republic

(3) Tartu Observatory, Faculty of Science and Technology, University of Tartu
Observatooriumi 1, Tõravere, 61602 Tartumaa, Estonia

(4) Astronomical Ring for Amateur Spectroscopy Group

**Abstract:** Z Andromedae is a prototype of classical symbiotic variable stars. It is characterized by alternating of quiescent and active stages, the later ones are accompanied by changes in both photometry and spectral characteristics of this object. The current activity of Z And began in 2000, and the last outburst was recorded at the turn of years 2017 and 2018. An important source of information about the behaviour of this symbiotic binary during the ongoing active stage is photometric and spectroscopic observations obtained with small telescopes by amateur astronomers. In this paper, we present the results of analysis of these observations, with an emphasis on the significant similarity of the last outburst of Z And with the previous ones, during which jets from this symbiotic system were observed. The presented results point to the importance of long-term monitoring of symbiotic binaries.

## Introduction

Symbiotic stars are strongly interacting systems, in which physical mechanisms related to transfer and accretion of matter cause observable activity. During their active stages, which may last from a few days to decades, they manifest increases of brightness (about 2–5 mag) and significant changes in their spectra. Usually, these systems consist of a cool giant of spectral type K–M and hot compact star, mostly a white dwarf. The mass transfer most likely takes place by the stellar wind of the cool giant, which is also the source of a dense circumbinary envelope. Symbiotic systems are detached binaries with orbital periods of hundreds to thousands of days.

Z Andromedae is a prototype of classical symbiotic binaries. The binary consists of a late-type M4.5 III giant (Skopal, 2008) and a white dwarf with a strong magnetic field, $B_S \geq 10^5$G (Sokoloski & Bildsten, 1999) accreting from the giant's wind. The orbital period of the binary system is 758 days (Mikołajewska & Kenyon, 1996). Sokoloski et al. (2006) proposed a combination of dwarf nova-like accretion disk instability and nova-like nuclear shell burning as an explanation for its outbursts. Distance estimates of Z And range from 0.6 to 2.19 kpc (average value $1.2 \pm 0.5$ kpc; Sokoloski et al., 2006). Recently, the parallax of $(0.512 \pm 0.030)$ mas was published for Z And in the Gaia DR2 (Gaia Collaboration et al., 2018) from which Bailer-Jones et al. (2018) derived the point distance of 1.84 kpc with uncertainty of (1.75–1.95) kpc representing 68% confidence interval.

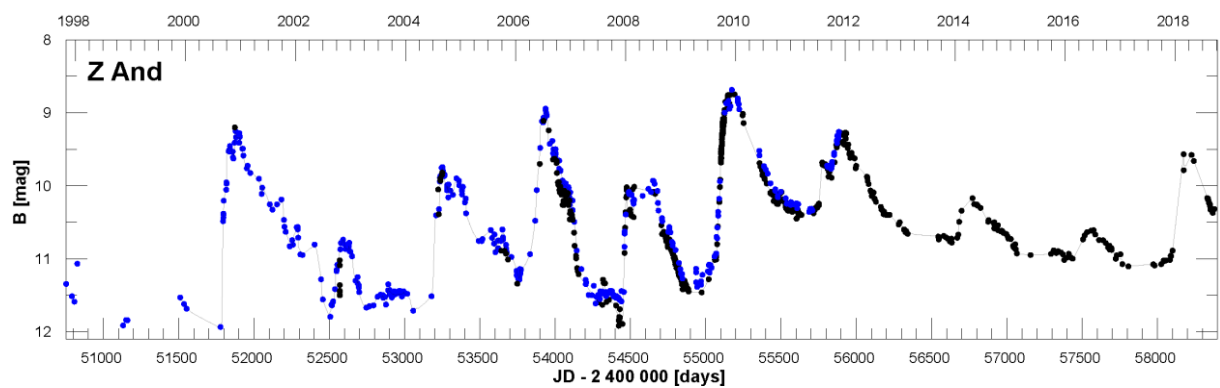

**Figure 1:** The light curve of Z And in the *B* filter over the period 1997–2018. Data from AAVSO and Skopal et al. (2002, 2004, 2007, 2012) are depicted in black and blue, respectively.

Z And is also one of a few known symbiotic stars producing collimated outflows or jets (Brocksopp et al., 2004; Leedjärv, 2004). The jets were detected in the radio images obtained during the 2000–2002 outburst (Brocksopp et al., 2004). The jet signatures were also recognised in the optical spectra of Z And acquired during the maxima of outbursts in 2006 (Skopal & Pribulla, 2006; Burmeister & Leedjärv, 2007; Tomov et al., 2007; Skopal et al., 2009) and 2009–2010 (Skopal et al., 2018).

During more than a hundred years of monitoring, Z And manifested several active stages with brightness variations ranging from a few tenths of a magnitude to about three magnitudes (Formiggini & Leibowitz, 1994; Skopal, 2008). The symbiotic system is in the active stage since the year 2000 (fig. 1). In this paper, we present the results of analysis of the spectroscopic observations obtained during the latest outburst of Z And which occurred at the beginning of the year 2018.

**Observations and analysis**

We have used medium resolution spectra of symbiotic star Z And obtained by the *ARAS* (Astronomical Ring for Amateur Spectroscopy) Group observers to study the activity and overall behaviour of Z And during the ongoing active stage. *ARAS*[3] is an initiative dedicated to promotion of amateur astronomical spectroscopy and professional/amateur collaborations. The observations of the group focus mainly on novae and symbiotic binaries. The network consists of independent amateur astronomers with small telescopes (20 to 60 cm) and spectrographs of different resolution (500 to 15000) covering the range from 3500 to nearly 8000 Å.

In the present study, we have analysed 61 spectra obtained by 6 observers from the overall number of 127 spectra (including low-resolution spectra that were not used) available in the *ARAS* database. In addition, we have used the spectra acquired on Tartu Observatory (Burmeister & Leedjärv, 2007) and Ondřejov Observatory (Skopal et al. 2009, 2018) to compare the latest outburst of Z And with the previous ones.

Our analysis was focused on the prominent emission lines in optical spectra of Z And: the neutral He I lines at λ 4471 Å and 6678 Å, ionized He II line at λ 4686 Å, the hydrogen lines Hβ at λ 4861 Å and Hα at λ 6563 Å. We measured equivalent widths (EWs) of these lines using the software *PlotSpectra* and computed the fluxes in lines using the photometric observations of Z And obtained from AAVSO International Database (Kafka, 2018) and from papers by Skopal et al. (2002, 2004, 2007, 2012).

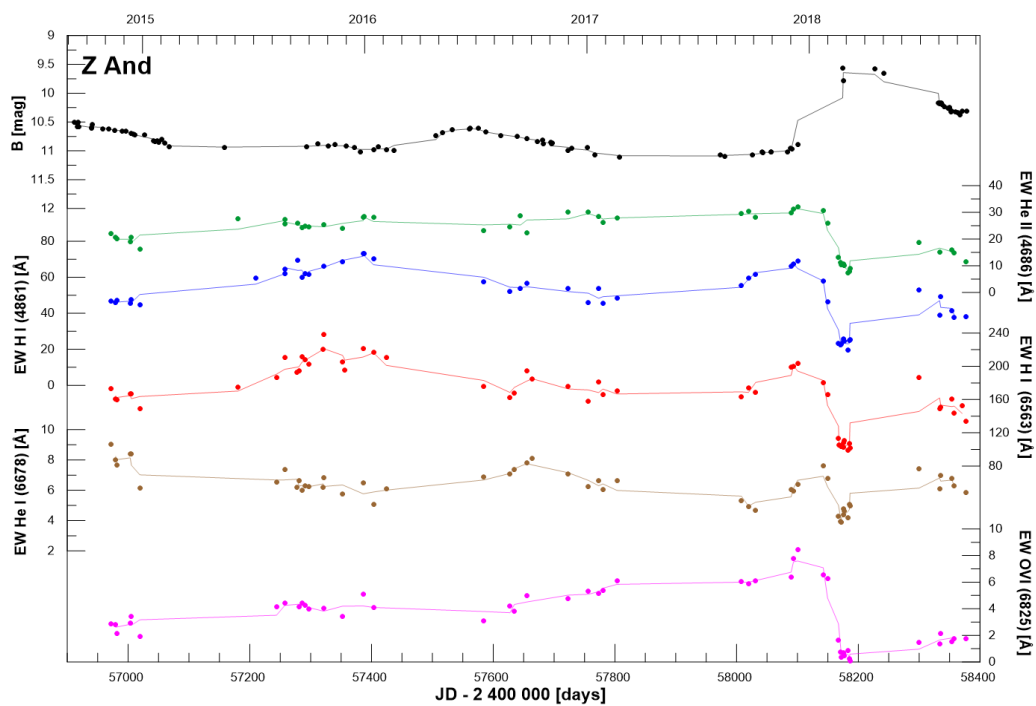

**Figure 2:** The EWs of prominent emission lines in spectra of Z And together with the AAVSO light curve in the *B* filter. The solid lines represent a spline fits to the data points.

---

[3] http://www.astrosurf.com/aras/Aras_DataBase/DataBase.htm

## Recent outburst activity of Z And

Since September 2000, the symbiotic system has manifested a typical Z And-type outburst activity with multiple outbursts of which the major ones occurred in December 2000, July 2006, December 2009 and December 2011. The recent outburst was observed at the beginning of the year 2018, and since then the brightness slowly declines (fig. 1). Photometrically, it was the most prominent outburst since 2012 with the maximal magnitudes of 9.6 and 9.0 in the *B* and *V* filters, respectively.

At the same period, the EWs of studied emission lines in optical spectra of Z And showed significant decline anticorrelated with brightness changes (fig. 2). In the case of AG Dra, such behaviour is typical for its *cool* outbursts and is related to the decrease of ionization source's temperature (Leedjärv et al., 2016). Moreover, the Raman-scattered O VI lines completely disappeared during the outburst. Other high ionisation lines such as [Fe VII] also disappeared during the outburst, but these lines reappeared at the time when O VI lines remained undetectable. The O VI lines are discussed in more details in separate section.

Similar behaviour of the emission lines in optical spectra of Z And was observed during some of its previous outbursts. In 2006, the symbiotic system underwent the strong outburst accompanied by the ejection of bipolar jets (Burmeister & Leedjärv, 2007; Tomov et al., 2007; Skopal et al., 2009). During that outburst, the He II line also practically disappeared. Despite the similarity of these two outbursts (in 2006 and 2018), no sign of the jet components around the Hα and Hβ lines was observed during the recent one (fig. 3). Their transient character is also supported by other observations: the jet components were detected during the outburst in 2008 and 2009–2010, but not during those in 2012 or 2014 (Skopal et al., 2018).

## Temperature evolution of the white dwarf

The behaviour of prominent emission lines in optical spectra of Z And suggests that the latest outburst of this symbiotic system could be of a *cool* type. We investigated the temperature evolution of the white dwarf in Z And using the method proposed by Iijima (1981) and modified by Sokoloski et al. (2006). Using some simplifications, which are discussed in detail in our recent paper (Merc et al., 2018), we can consider the He II/Hβ ratio as a proxy to the temperature of the hot component in symbiotic binaries:

$$T_{hot}(10^4 K) \approx 14.16 \sqrt{\frac{EW_{4686}}{EW_{H_\beta}}} + 5.13$$

Even though the applied simplifications allow to obtain only the upper limit of the temperature (Merc et al., 2018), this method can still be used to study the temperature changes of the hot component of symbiotic systems. Applying this method to Z And, we confirmed the decrease of the white dwarf's temperature during and after the last outburst of this symbiotic system (Fig. 4). During the period 2014–2018, the average value of the ratio was 0.44 which is corresponding to the white dwarf's temperature around 145 000 K. In the course of the outburst 2018, its value dropped to about 0.29 (about 129 000 K).

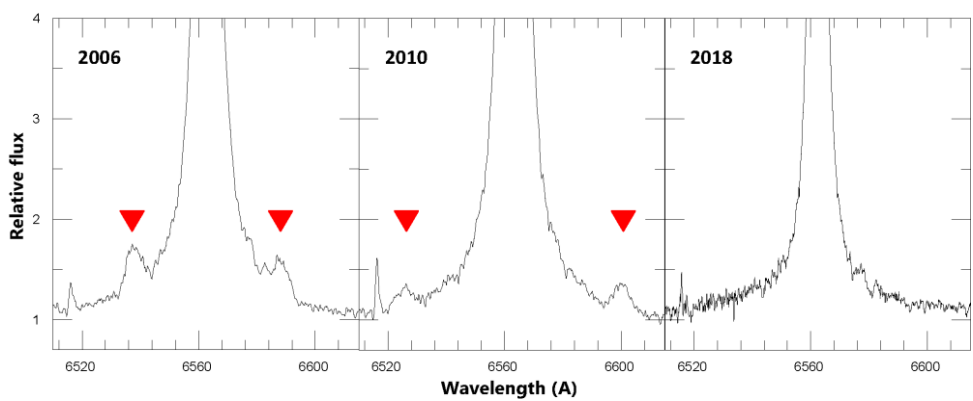

**Figure 3:** The jet components (marked by the red triangles) around the Hα emission line occurred during the outbursts of Z And in 2006 and 2010. No sign of the jet components was observed during the recent outburst in 2018. The spectra from 2006 and 2010 were obtained on Ondřejov Observatory, the spectrum from 2018 is from the *ARAS* database.

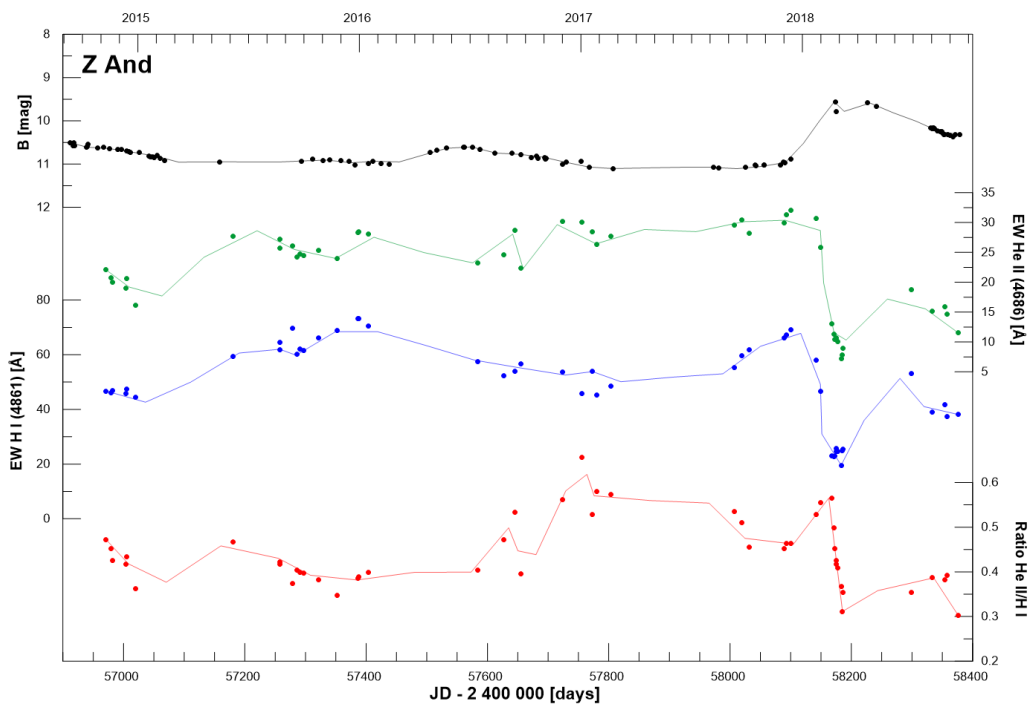

**Figure 4:** The EW ratio of the two strong emission lines He II and Hβ together with their EWs and AAVSO light curve in the *B* filter. The solid lines represent a spline fits to the data points.

## The Raman-scattered O VI lines

The Raman-scattered O VI lines are broad emission features in optical spectra at 6825 and 7082 Å which are a product of Raman-scattering of the photons of the O VI resonance lines at 1032 and 1038 Å off the atoms of neutral hydrogen (Schmid, 1989). They occur almost exclusively in the spectra of symbiotic stars. During recent outburst of Z And, the Raman-scattered O VI line at 6825 Å disappeared due to cooling of the ionizing source (fig. 2 and 4). Similar vanishing was observed during the outburst of AG Dra in 2006, confirming a drop in the white dwarf's temperature (Leedjärv et al., 2016).

It is worth to note, that similar disappearance of the O VI lines was also observed during the recent *hot* outbursts of AG Dra and AG Peg (Skopal et al., 2017; Merc et al., 2019) when the temperature of the ionizing source in these symbiotic systems has risen. In these cases, vanishing of the O VI lines could be probably caused by an increase of mass-loss rate from the hot component which makes O VI zone optically thick (Skopal et al., 2017). It is possible that a similar effect also played a role in the recent outburst of Z And because the O VI lines were undetectable long after other highly ionized lines reappeared in the spectrum. On the other hand, according to the new *ARAS* observations, during the current post-outburst decline of the new symbiotic star HBHa 1704-05 (Munari et al., 2018), the Raman-scattered O VI lines have appeared, while the [Fe VII] lines were still very weak.

## Conclusions

In the present paper, we investigated the latest outburst of symbiotic binary Z And. We analysed photometric and spectroscopic observations covering this brightening with focus on the evolution of the EWs of selected prominent emission lines and their profile changes. As we have shown, the recent outburst manifested the decrease of the emission line EWs and the steep decrease of the ionising source temperature. Similar behaviour of Z And was observed in the past, especially during the major outburst in 2006 which was accompanied by the ejection of bipolar jets, signatures of which were observed in the profile of hydrogen lines. Despite the significant similarity between the outbursts in 2006 and 2018, no jet components were observed in the spectra during the latter one. Nevertheless, it is worth observing this interesting symbiotic system further. Only long-term monitoring can help to uncover the physical mechanisms responsible for the unusually long active stage of Z And lasting now for almost two decades.

Moreover, the presented results showed also the importance of professional/amateur collaborations. *ARAS* Group is a perfect example that such collaboration can be very successful and can bring important results. Thanks to amateur photometric and spectroscopic data, we are now able to monitor the evolution of symbiotic systems on timescales which were not previously available.

**Acknowledgement**

We are grateful to all *ARAS* members that contributed their observations to this paper, particularly we acknowledge and thank Paolo Berardi, Joan Guarro Flo, Tim Lester, Jacques Montier and Peter Somogyi. We acknowledge with thanks the variable star observations from the AAVSO International Database contributed by observers worldwide and used in this research. This work was supported by the Slovak Research and Development Agency grant No. APVV-15-0458 as well as by the Estonian Ministry of Education and Research institutional research funding grant IUT 40-1.